\documentclass[12pt]{article}
\usepackage{amssymb}
\usepackage{amsbsy}

\def\nextline{\hfill\break}
\def\mycomm#1{\nextline\strut\kern-3em{\tt ====> #1}\nextline}
\def\VV{{\hbox{$\boldsymbol{v}$}}} 
\def\SS{{\hbox{$\boldsymbol{s}$}}} 

\def\gray{\special{ps: 0.5 setgray}}
\def\black{\special{ps: 0.0 setgray}}

\newcommand{\draft}{
\newcount\timecount
\newcount\hours \newcount\minutes  \newcount\temp \newcount\pmhours
\hours = \time
\divide\hours by 60
\temp = \hours
\multiply\temp by 60
\minutes = \time
\advance\minutes by -\temp
\def\hour{\the\hours}
\def\minute{\ifnum\minutes<10 0\the\minutes
            \else\the\minutes\fi}
\def\clock{
\ifnum\hours=0 12:\minute\ AM
\else\ifnum\hours<12 \hour:\minute\ AM
      \else\ifnum\hours=12 12:\minute\ PM
            \else\ifnum\hours>12
                 \pmhours=\hours
                 \advance\pmhours by -12
                 \the\pmhours:\minute\ PM
                 \fi
            \fi
      \fi
\fi
}
\def\fullclock{\hour:\minute}
\begin{centering}
\gray
\special{ps: -90 rotate}
\special{ps: -4600 -3800 translate}
\font\Hugett  =cmtt12 scaled\magstep4
\hbox{\Hugett Draft: \today, \clock}
\black
\special{ps: 90 rotate}
\special{ps: 3800 -4600 translate}
\end{centering}
\vskip -1.7cm
$\phantom{a}$
} 

\newcommand{\bmath}{\begin{displaymath}}
\newcommand{\emath}{\end{displaymath}}

\def\beq{\begin{equation}}
\def\eeq{\end{equation}}

\newcommand{\bea}{\begin{eqnarray}}
\newcommand{\eea}{\end{eqnarray}}

\def\eqref#1{(\ref{#1})}
\def\bra#1{\left\langle #1\right|}
\def\ket#1{\left| #1\right\rangle}
\def\sinxi{$ \sin \xi $}

\newcounter{saveeqn}

\catcode`\@=11 

\def\eqarraylabel#1{\@bsphack \if@filesw
{\let \thepage \relax \def \protect {\noexpand \noexpand \noexpand }%
\edef \@tempa {\write \@auxout {\string \newlabel
{#1}{{\mbox{\arabic{saveeqn}}}{\thepage }}}}\expandafter }\@tempa
\if@nobreak \ifvmode \nobreak \fi \fi \fi \@esphack}

\catcode`\@=12 

\def\tsize{\Large} 
\def\asize{\normalsize} 
\title {
\begin{flushright}
$\phantom{a}$\\
\vspace{-4cm}
\normalsize TAUP-2694-02\\
\normalsize WIS/3/02-JAN-DPP\\
\end{flushright}
\vspace{1.0cm}
\tsize
On Reconciling Gottfried Sum Rule Violation\\
\tsize
with Cabibbo Theory\thanks{Supported
in part by grant from US-Israel Bi-National Science Foundation}
}
\author{\asize
Marek Karliner$\,^{1}$~\footnote{\tt e-mail: marek@proton.tau.ac.il}\\
\asize and\\
\asize $\phantom{a}$ Harry J. Lipkin$\,^{1,2}$~\footnote{\tt e-mail:
harry.lipkin@weizmann.ac.il}
\vspace{0.5cm}
\\
\asize \sl $^1$\,School of Physics and Astronomy\\
\asize \sl The Raymond and Beverly Sackler Faculty of Exact Sciences\\
\asize \sl Tel Aviv University, 69978 Tel Aviv, Israel\\
\asize \sl and\\
\asize \sl $^2$\,Department of Particle Physics\\
\asize \sl The Weizmann Institute of Science, 76100 Rehovot, Israel\\
}
\date { }
\begin {document}
\maketitle
\maketitle
\begin{abstract}
We discuss the seemingly contradictory constraints of simultaneously
preserving the $SU(3)$-symmetric 
Cabibbo description of the weak vector baryon matrix
elements,
accounting for $SU(3)$ flavor symmetry breaking and
describing the observed violation of the Gottfried Sum Rule.
 We try to construct a simple model that will satisfy these
constraints and use it to explain the generic difficulties and tradeoffs.
    \end{abstract}
\thispagestyle{empty} 
%
\newpage
\section{Introduction}

So far there has been no successful model for the flavor structure of the
proton which is simultaneously consistent with three established
experimental results:
\begin{enumerate}
\item
The agreement of the data on weak semileptonic baryon decays produced
by the conserved vector current with the predictions of Cabibbo theory.
 The vector matrix
element is uniquely determined by Cabibbo theory in the $SU(3)$ symmetry
limit.
\item
The necessity for flavor $SU(3)$ breaking implied by the $K-\pi$
mass difference and the mass difference between strange and nonstrange
quarks and confirmed experimentally by the observation of a suppression
of the strange component in the sea of $q \bar q$ pairs in the nucleon.
 The strange quark
contribution to the proton sea is already known from experiment
to be reduced roughly by a factor of two from that of a flavor-symmetric
sea \cite{CCFR}.
\item
The observed violation of the Gottfried sum rule (see
 \cite{Kumano:1998cy} for a review).
This requires an isovector component in the sea. Isospin invariance then
immediately dictates that the proton wave function contains a valence
neutron and a positively charged sea ($u\bar d$). A well-known implementation
of such a fluctuation is $ p \leftrightarrow n \pi^+$.
Had $SU(3)$ been unbroken, it would in turn require, in analogy with the 
isovector invariance, that the proton wave function contains 
a valence hyperon and a sea with net strangeness, like in 
$ p \leftrightarrow \Lambda K^+$. There is little experimental 
support for the presence of such a component in the proton wave function.
We thus see that the violation of the Gottfried sum rule provides 
additional substantial evidence of a significant $SU(3)$ breaking in the 
nucleon wave function.
\end{enumerate}

There are models which are consistent with Cabibbo theory and $SU(3)$
breaking in the sea by assuming that only valence quarks contribute to
hyperon decays and that the $SU(3)$  breaking occurs only in the sea and
is the same for all baryons.  One such  model \cite{PJECH} for breaking
$SU(3)$ via the mechanism (2) keeps all the good results of Cabibbo theory
by introducing a flavor asymmetric sea with no net flavor quantum numbers
into a baryon wave function whose valence quarks satisfy $SU(6)$ symmetry
and whose sea is {\em the same} for all baryons.  In ref. \cite{Karlip}
it was shown that in this model all charged current
matrix elements are given by the valence quarks. This provides an explicit
justification for the hand-waving argument \cite{PJECH} in a proposed
toy model that in the hyperon decay the sea behaves as a spectator. In
particular, for the strangeness-changing vector charge producing the
$\Sigma^- \rightarrow n$ decay. But such models fail to account for the
violation of the Gottfried sum rule.

Other models introduce the violation of the Gottfried sum rule by
introducing a sea which is not isoscalar. These can incorporate the observed
strange quark suppression in the sea, but so far only at the price of violating
the predictions of Cabibbo theory for hyperon decays. They can be made
consistent with Cabibbo theory by retaining full $SU(3)$ symmetry. But this
cannot be consistent with the observed $SU(3)$ breaking effects. For example,
if one postulates
a pion cloud around the nucleon to explain the violation of the
Gottfried sum rule, one is required by $SU(3)$ symmetry to
also include a kaon cloud
around a valence hyperon in the same wave function, and cannot explain the
observed suppression of strange quarks in the sea.

We thus see that in any model which includes a pion cloud in
the proton wave function, $SU(3)$ breaking must reduce the kaon cloud
relative to the pion cloud from the value in the symmetry limit. This
necessary
breaking seems to have a serious effect on the matrix elements of the
strangeness-changing current responsible for hyperon decays.
The nature of this inconsistency is illuminated by noting that production
of a state of strangeness +1 by the action of the $SU(3)$ generator $V_+$
($s \rightarrow u$ and $\bar u \rightarrow \bar s$) when acting on a
``nucleon+pion" 
proton model wave function, indicates that this proton wave function is
not a pure $SU(3)$ octet but contains a {\bf 27} admixture: 
\beq
\ket{N\pi}  \ni \ket{p \pi^0 }:\qquad\qquad
V_+ \ket{p\pi^0 }= \ldots + \ket{p K^+ }
\label{VplusOnp}
\eeq
When $SU(3)$
symmetry is restored in this model wave function by adding the correct
admixture of $\Lambda \,K$, $\Sigma\, K$ and $p \,\eta_8 $ states to the
wavefunction with the pion cloud,  the action of the operator $V_+$ on
these components produces the $p\,K^+$ state with just the right phase
to cancel the $p\,K^+$ state produced on the nucleon-pion state.

We now wish to generalize the treatment for the case of the pion cloud to the
general case of a  physical proton wave function with a valence nucleon and a
sea of quark-antiquark pairs in which the numbers of $\bar u$ and $\bar d$ are
different and the difference is adjusted to fit the violation of the Gottfried
sum rule.
However, we wish to preserve isospin symmetry, which
requires that a proton wave function with a valence proton  and a sea that is
not isoscalar must also have a component with a valence neutron and a sea which
carries a positive electric charge to give the electric charge of the physical
proton. We therefore include such components in our proton  wave function.
Similarly flavor $SU(3)$ symmetry would require including components with a
valence hyperon and with an excess of strange $\bar s$ antiquarks in the sea to
balance the  valence strangeness of the hyperon. We do not wish to include such
components  because $SU(3)$ is experimentally known to be broken, and there is a
deficiency of strange $\bar s$ antiquarks in the sea and not an excess. We
therefore assume that our baryon wave functions have a sea which can contain $s
\bar s$ pairs but no net strangeness; e.g they can contain pion clouds but no
kaon clouds.

We shall now show that such baryon wave functions are not consistent with  the
predictions of Cabibbo theory for hyperon decays.
The transition matrix
elements for the weak vector semileptonic decays of the $\Lambda$ and
$\Sigma^o$ hyperons  to this proton wave function cannot give the results
predicted by  Cabibbo theory and which agree with  experiment.
%
%

Let us write the general wave function for the physical proton as
\beq
\ket{p_{phys}} = \cos \xi\cdot
  \ket {p_{val}\cdot \SS}  +
  {{\sin \xi}\over{\sqrt 3}} \cdot
  [\ket {p_{val}\cdot \VV_o}  -
{\sqrt 2}\ket {n_{val}\cdot \VV_+}]
\label{physpgen}
\eeq
where $p_{val}$ and $n_{val}$ denote respectively valence proton and neutron,
$\SS$ denotes an isoscalar sea of $q \bar q$ pairs and $\VV_o$ and 
$\VV_\pm$ denote
the three components of an isovector sea.  The parameter $\xi$ determines
the isospin asymmetry in the nucleon sea.

A basic ambiguity arises in any formulation which separates quarks into valence
and sea, as there is no physical label on a given quark to specify whether it is
a valence quark or a sea quark. We assume here that the valence and sea quarks
have very different momentum distributions, with the valence quarks being
``hard" and the sea quarks ``soft", and that the overlap region between the two
momentum distributions is negligible. This classification can break
down in matrix elements describing processes with high momentum transfer, where
an initial quark with soft momentum can turn into a final quark with hard
momentum and vice versa. However we are concerned here only with matrix elements
having essentially zero momentum transfer and only require that the overlap
region between valence and sea quark momentum distributions be negligibly small.  

For a specific example we can write
\bea
\ket{p_{phys}} &=& {\cos \xi\over\sqrt 2}\cdot
  \left[ \,\ket {p_o\cdot (d \bar d)_\SS}
+ \ket {p_o\cdot (u \bar u)_\SS}\,\right] +
\nonumber
\\
  &+&{\sin \xi\over\sqrt 6} \cdot
\left[\,\ket {p_o\cdot (d \bar d)_\VV} - \ket {p_o\cdot (u \bar u)_\VV} -
2\ket {n_o\cdot (u \bar d)_\VV}\,\right]
\label{physp}
\eea
where $p_o$ denotes a valence proton surrounded by an isoscalar sea of  $q \bar
q$ pairs which are inert and do not contribute to the violation of the
Gottfried sum rule, nor to the weak strangeness-changing transitions,  and
similarly $n_o$ denotes a valence neutron surrounded by an isoscalar sea. The
additional isoscalar and isovector $q \bar q$ pairs are expressed explicitly in
states labeled by the subscripts $\SS$ and $\VV$.

The difference between the  numbers
of $\bar d$ and $\bar u$ antiquarks in the proton is given by
\bea
\bra{p_{phys}}\delta N(\bar q) \ket{p_{phys}} &=&
 {{\sin 2\xi}\over{\sqrt 3}}\cdot
Re \bra{p_{val}\cdot \SS}\delta N(\bar q)\ket {p_{val}\cdot \VV_o}
+\nonumber\\
&+& {2\over 3}\cdot  \sin ^2 \xi \cdot
\bra{n_{val}\cdot \VV_+}\delta N(\bar q)\ket {n_{val}\cdot \VV_+}
\eea
where $\delta N(\bar q) \equiv N(\bar d) - N(\bar u) $.
For the specific example (\ref{physp})
\beq
N(\bar d) - N(\bar u) =  {{\sin 2\xi\,\cos \phi}\over{\sqrt 6}}
+ {2\over 3}\cdot  \sin ^2 \xi
\eeq
where $\cos \phi$ denotes the overlap between the isoscalar and isovector
states,
\beq
\cos \phi \equiv  \langle {p_o\cdot (d \bar d)_\SS}
  \ket {p_o\cdot (d \bar d)_\VV}
\label{overlap}
\eeq

We note here that the difference between the number of $\bar d$ and $\bar u$
antiquarks in the proton has two contributions, one linear in the isospin
asymmetry parameter \sinxi\ and one quadratic. The linear term may be
crucial in allowing a small value of $\sin \xi$ to produce a violation of the
Gottfried sum rule, while the violations of Cabibbo theory will be 
shown below to be quadratic in
$\sin \xi$ and can therefore be much smaller.  However, the linear term depends upon
the overlap between an isoscalar sea and an isovector sea. This overlap
vanishes if the isovector sea is due to a pion cloud, since there is no
isoscalar partner to the pion. Therefore the relative magnitudes of the 
violations of the Gottfried sum rule and Cabibbo theory are model dependent
and depend upon the relative magnitudes of the linear and quadratic terms.  

An example of a model which is probably not realistic and would give such a
linear term has a sea due to a vector meson cloud, rather than to a pion
cloud. Here the isoscalar sea is due to the $\omega$ and the isovector to the
$\rho$. Interference between the $\rho$ and $\omega$ amplitudes can produce the
linear term that violates the Gottfried sum rule.

We now examine the symmetry properties of this wave function to check that it
indeed satisfies isospin symmetry and whether its manifest $SU(3)$ symmetry
breaking must necessarily lead to violation of the predictions of Cabibbo
theory.

We first construct the wave function for the physical neutron and check
that these wave functions are a doublet of isospin 1/2 by applying the
isospin raising and lowering operators $I^\pm$ to these wave functions,
\beq
\ket{n_{phys}} = I^- \ket{p_{phys}} =
 \cos \xi\cdot
  \ket {n_{val}\cdot \SS}  +
  {{\sin \xi}\over{\sqrt 3}} \cdot
  [\ket {n_{val}\cdot \VV_o}  -
{\sqrt 2}\ket {p_{val}\cdot \VV_-}]
\label{physngen}
\eeq
For the specific example (\ref{physp}),
\bea
\ket{n_{phys}} &=& {\cos \xi \over\sqrt 2}  \cdot
\left[\,\ket{n_o\cdot (d \bar d)_\SS} 
+ \ket {n_o\cdot (u \bar u)_\SS}\,\right] +
\nonumber\\
&+&{\sin \xi\over\sqrt 6} \cdot
\left[\,\ket {n_o\cdot (u \bar u)_\VV} -  \ket {n_o\cdot (d \bar d)_\VV} -
2 \ket {p_o\cdot (d \bar u)_\VV}\,\right]
\label{physn}
\eea
\beq
I^+ \ket{p_{phys}} =  {{\sin \xi}\over{\sqrt 6}} \cdot
 2[\ket {p_o\cdot (u \bar d)_\VV} -
\ket {p_o\cdot (u \bar d)_\VV}] = 0
\eeq
\beq
I^- \ket{n_{phys}} =  {{\sin \xi}\over{\sqrt 6}} \cdot
 2[\ket {n_o\cdot (d \bar u)_\VV} -
\ket {n_o\cdot (d \bar u)_\VV}] = 0
\eeq
These wave functions satisfy the constraints of isospin, and the parameter $\xi$
can be fixed to satisfy the Gottfried sum rule.

We now investigate the action of the strangeness-changing components of the
charged weak vector current on the proton wave function (\ref{physp}).
At zero-momentum transfer, these are just the $V\!$-spin raising and lowering
operators, denoted by $V_\pm$, which generate
$u \,\lower0.15em\hbox{$\leftrightarrows$}\, s$
and
$\bar s \,\lower0.15em\hbox{$\rightleftarrows$}\, \bar u$ transitions at
the quark level \cite{Karlip}.  The requirement that the proton and
$\Lambda$ are members of the same $SU(3)$ octet gives the two conditions:
\beq
V_+ \ket {p_{phys}} = 0
\label{V1}
\eeq
\beq
P(I=0) \cdot V_- \ket {p_{phys}} = {{\sqrt 6}\over{2}} \ket {\Lambda_{phys}}
\label{V2}
\eeq
where $P(I=0)$ denotes a projection operator which projects out the $I=0$
component of the wave function and $\ket {\Lambda_{phys}} $ denotes the
normalized physical $\Lambda$ wave function. These two conditions required by
Cabibbo theory were shown in ref.~\cite{Karlip} to be manifestly violated by
the proton wave function with a pion cloud.

The condition (\ref{V1}) is also seen to be violated by the physical proton
wave function (\ref{physp}) used here, since
\beq
V_+ \ket {u \bar u} = - \ket {u \bar s}; ~ ~ ~
V_+ \ket {s \bar s} = \ket {u \bar s}
\label{Vpair}
\eeq

When the sea is $SU(3)$ symmetric, the numbers of ${u \bar u}$ and  ${s \bar s}$
pairs are equal and the condition (\ref{V1}) is satisfied because the two terms
in eq. (\ref{Vpair}) cancel.  This cancellation no longer occurs when the
numbers of ${u \bar u}$ and  ${s \bar s}$ are unequal, as is required to fit
the known suppression of the strange component in the nucleon
sea~\cite{CCFR}.

We now attempt to bypass the $SU(3)$ breaking effects indicated by the violation
of the condition (\ref{V1}) by  constructing baryon wave functions such that
the transition matrix elements between the physical  baryon states satisfy
$SU(3)$, even though the physical baryon wave functions are no longer members
of the same $SU(3)$ octet; i.e. they contain admixtures of other representations.

We assume that all the baryon wave functions are constructed like
$\ket{p_{phys}}$ having no valence hyperons and a sea which can contain
$s \bar s$ pairs but no net strangeness; e.g they can contain pion clouds but no
kaon clouds.

In order to examine closer the action of $V_{\pm}$ on the sea, we now
define modified $V\!$-spin raising and lowering
operators  $V_+^{val}$ and $V_-^{val}$, which by definition
act only on the valence baryon
component of the wave function and not on the sea quarks, therefore
leaving the sea unchanged.

We now  note that the operation of the strangeness-changing operators on a sea
with no net strangeness creates a sea with nonvanishing strangeness. Since our
baryon wave functions are constructed to have no net strangeness 
in the sea,
\beq
\bra{\Lambda_{phys}}[V_- - V_-^{val}] \ket{p_{phys}} =
\bra{\Sigma^o_{phys}}[V_- - V_-^{val}] \ket{p_{phys}} = 0. \label{valeq} \eeq

This is seen in  specific example where
\beq
[V_-  - V_-^{val}]
\ket{p_{phys}} = 
{\cos \xi\over\sqrt 2}\cdot   
\ket {p_o\cdot (s \bar u)_\SS}+ {{\sin \xi}\over{\sqrt 6}} \cdot
 [\ket {p_o\cdot (s \bar u)_\VV} -
2\ket {n_o\cdot (s \bar d)_\VV}]
\label{seaflavor}
\eeq
The RHS of eq (\ref{seaflavor}) has a sea with nonvanishing strangeness.

Here our assumption neglecting the overlap region between the valence and sea
quark momentum distributions is seen to be essential in any model which
attempts to preserve the SU(3) relations of Cabibbo theory despite the large
SU(3) breaking in the sea. The operator $V_-$ acting on the  proton sea can
create a strange quark in a final hyperon, both via the ${u \bar u}
\rightarrow  {s \bar u}$ and the ${s \bar s}\rightarrow  {s \bar u}$
transitions. Since the ${s \bar s}$ component in the proton sea is suppressed
by SU(3) breaking, any contribution from these transitions to the
matrix elements between physical nucleon and hyperon states will violate SU(3)
in the physical matrix elements and violate Cabibbo theory. Thus it is necessary
to assume that the SU(3) breaking remains in the sea and affects only the
magnitude of a sea with nonvanishing strangeness which has no overlap with the
baryon wave functions. 
Thus only valence quarks contribute to hyperon decays
as in a previous toy model\cite{PJECH}.

Under this assumption
the transition matrix elements for the $\Lambda \rightarrow p$ and
$\Sigma^o \rightarrow p$ decays are respectively
$\bra {p}V_+^{val} \ket{\Lambda}$ and $\bra {p}V_+^{val}\ket{\Sigma^o} $.
We now note that
\beq
|\bra {p}V_+^{val} \ket{\Lambda}|^2 + |\bra {p}V_+^{val}\ket{\Sigma^o}|^2
\leq \sum_i  |\bra {p}V_+^{val} \ket{i}|^2 =
\bra {p}V_+^{val} V_-^{val} \ket{p}
\label{inequalityI}
\eeq
where the sum over $i$ denotes a complete set of states.

\noindent
Since $V_+^{val} \ket{p} = 0$, we can replace the product
$V_+^{val} V_-^{val}$
in (\ref{inequalityI}) by the commutator,
\beq
|\bra {p}V_+^{val} \ket{\Lambda}|^2 + |\bra {p}V_+^{val}\ket{\Sigma^o}|^2
\leq \bra {p}[V_+^{val}, V_-^{val}] \ket{p}
= 2\bra {p}V_z^{val} \ket{p}                                                     \eeq
Substituting the expression (\ref{physp})
for the physical proton wave function and
noting that the eigenvalues of $V_z^{val}$ for the proton and neutron are
respectively +1 and +(1/2), we obtain
\beq
|\bra {p}V_+^{val} \ket{\Lambda_o}|^2 + |\bra {p}V_+^{val}\ket{\Sigma^o}|^2
\leq 2\cos^2 \xi   +
{\textstyle {4\over 3}} \cdot \sin^2 \xi
= 2  -
{\textstyle{2\over 3}} \cdot \sin^2 \xi
\label{lambda_sigma}
\eeq
Thus the sum of the semileptonic vector decay rates for the $\Lambda
\rightarrow p$ and $\Sigma^o \rightarrow p$ decays agrees
with the
value 2 predicted by Cabibbo theory only when $\sin^2 \xi = 0$, i.e. when
there is no flavor asymmetry in the sea and the Gottfried sum rule is
not violated.

We have shown that explaining the observed
violation of the Gottfried sum rule while keeping the good
results of the Cabibbo theory requires the introduction of 
net strangeness in the nucleon sea. The latter seems to be in conflict with
experiment. There are 
two possible directions for avoiding this conflict.

\begin{enumerate}
\item
It may not be a real conflict with the real data. This requires a
quantitative analysis of how much violation of Cabibbo theory is allowed by the
real data with real errors. The question remains of whether violations of both
the Gottfried sum rule and Cabibbo theory can be consistent with present data.
In that case better data to reduce the errors will be of interest.
The answer to this question is model-dependent, since it depends upon the
overlap  (\ref{overlap}) between isoscalar and isovector seas.

\item
The second logical possibility, as unlikely as it may sound is that 
there may be a small component of 
``valence like" strange quarks in the proton. 
By ``valence like" we mean quarks with large values of $x$, usually 
associated with valence components of the nucleon wave function. Of course
the total number of strange plus anti-strange quarks remains zero, but the 
respective $x$ distributions might be very different.
This can be checked by
better measurements of the $x$-dependence of the strangeness in the proton. 
Indeed, there have been suggestions in the literature (see for example
\cite{Brodsky:1996hc}-\cite{Ma:2000uv}),
that the strange sea might exhibit a considerable degree of asymmetry.
However, as far as we are aware, until now there has been no discussion
of whether such an asymmetric strange sea is compatible with the successful
Cabibbo theory.
Here again quantitative limits are needed.
\end{enumerate}

\section*{Acknowledgments}
This research was supported in part by a grant from the
United States-Israel Binational Science Foundation (BSF), Jerusalem,
Israel, and by the Basic Research Foundation administered by the Israel
Academy of Sciences and Humanities.

\end {document}